\documentclass[manuscript]{aastex}

\shorttitle{Horizontal Divergent Flow prior to Flux Emergence}
\shortauthors{Toriumi et al.}

\begin{document}


\title{Detection of the Horizontal Divergent Flow
  prior to the Solar Flux Emergence}


\author{S. Toriumi$^{1}$, K. Hayashi$^{2}$, and T. Yokoyama$^{1}$}
\affil{$^{1}$Department of Earth and Planetary Science,
  University of Tokyo,
  7-3-1 Hongo, Bunkyo-ku, Tokyo
  113-0033}
\email{toriumi@eps.s.u-tokyo.ac.jp}
\affil{$^{2}$W. W. Hansen Experimental Physics Laboratory,
  Stanford University,
  Stanford, CA 94305, USA}



\begin{abstract}
  It is widely accepted
  that solar active regions
  including sunspots
  are formed
  by the emerging magnetic flux
  from the deep convection zone.
  In previous numerical simulations,
  we found that
  the horizontal divergent flow (HDF)
  occurs before the flux emergence
  at the photospheric height.
  This Paper reports
  the HDF detection
  prior to the flux emergence
  of NOAA AR 11081,
  which is located
  away from the disk center.
  We use SDO/HMI data
  to study the temporal changes
  of the Doppler and magnetic patterns
  from those of the reference quiet Sun.
  As a result,
  the HDF appearance
  is found to
  come before
  the flux emergence
  by about 100 minutes.
  Also,
  the horizontal speed
  of the HDF
  during this time gap
  is estimated to be
  0.6 to 1.5 km s$^{-1}$,
  up to 2.3 km s$^{-1}$.
  The HDF is caused
  by the plasma
  escaping horizontally
  from the rising magnetic flux.
  And the interval between
  the HDF and the flux emergence
  may reflect
  the latency
  during which
  the magnetic flux
  beneath the solar surface
  is waiting
  for the instability onset
  to the further emergence.
  Moreover,
  SMART H$\alpha$ images
  show that
  the chromospheric plages appear
  about 14 min later,
  located co-spatial with
  the photospheric pores.
  This indicates
  that the plages
  are caused by plasma
  flowing down along the magnetic fields
  that connect the pores
  at their footpoints.
  One importance
  of observing the HDF
  may be the possibility
  to predict
  the sunspot appearances
  that occur in several hours.
\end{abstract}





\section{Introduction\label{sec:introduction}}

Solar active regions (ARs)
including sunspots
are generally thought to be
the consequence of flux emergence,
that is, the buoyant rise
of the magnetic flux
from the deep convection zone
\citep{par55}.
Observationally,
the new emerging flux appears
as a small and bright bipolar plage
in the chromospheric \ion{Ca}{2} H and K line cores
\citep{fox08,she69}.
Soon afterwards,
the arch filament system (AFS)
composed of parallel dark fibrils
appears in the line core of H${\rm \alpha}$
\citep{bru67}.
The fibrils are magnetic field lines
connecting the faculae
of positive and negative polarities.
In the photosphere,
small pores are formed
at the root of chromospheric filaments
with downflows up to
$\sim 1\ {\rm km\ s}^{-1}$.
The faculae of opposite polarity
separates,
initially at the rate of $>2\ {\rm km\ s}^{-1}$,
and the rate drops
to $1.3$--$0.7\ {\rm km\ s}^{-1}$
during the next 6 hours \citep{har73}.
New magnetic flux emerges continuously
within the opposite polarities.
If the flux is sufficient,
the pores are gathered,
and gradually sunspots are formed
near the leading and the following plages
\citep{zir72}.

In the last several decades,
numerical computations have been
well developed
to reveal the dynamics
of the flux emergence
and the birth of the active region
\citep[e.g.][]{shi89}.
In our recent simulations
on the large-scale flux emergence
from a depth of $20,000\ {\rm km}$,
the rising twisted flux tube
in the convection zone
decelerates and makes
a flat structure
just beneath the photosphere
\citep[e.g.][]{tor12}.
In this calculation,
the plasma,
which is pushed up
by the rising flux,
escapes laterally
around the surface.
The appearance
of the divergent outflow
at the photosphere
was found to be earlier than
that of magnetic flux,
and, at this moment,
the outflow is mainly horizontal.
Hereafter
we call this preceding outflow
as a horizontal divergent flow (HDF).
A similar flow
is also reported
by \citet{che10}.
However,
to our knowledge,
the HDF
prior to the flux emergence
has not been confirmed clearly
in previous observations
\citep{kos09}.
Here, we use 
the term ``horizontal''
to indicate the direction
parallel to the solar surface.

The aim of this study
is to investigate the HDF
and the evolving magnetic field
at an early phase
of the flux emergence.
For this purpose,
we used
the Dopplergrams and magnetograms
of the Helioseismic and Magnetic Imager (HMI)
on board the Solar Dynamics Observatory (SDO),
since their continuous observations
of the whole solar disk
make it possible
to achieve information
at the very moment of,
or even before the flux emergence
at the surface.

Our numerical result
indicates that,
if the newly emerging region
is located away from the disk center,
if a pair of
positive and negative Doppler patterns
is detected
just before the flux emergence,
and if the positive (negative) pattern
is limbward (disk-centerward),
the observed Doppler velocity
is mainly horizontal
rather than vertical.
Therefore,
we can evaluate the horizontal velocity
of the escaping plasma
from the Doppler velocity,
by considering the heliocentric angle
of the active region
from the disk center.
One advantage of this method
over the ordinal local correlation-tracking method
\citep{nov88}
is that
the horizontal velocity
of the plasma
can be evaluated independently
of the apparent motion
of magnetic elements
at the photosphere.
After the flux has emerged,
we can not obtain
the horizontal speed
from the Doppler velocity,
since it may contain a vertical motion
such as rising of magnetic fields
or a downflow
in the convective collapse process.

In this Paper,
we report the first determination
of the HDF
prior to the flux appearance,
using SDO/HMI Dopplergrams and magnetograms.
We also studied
the chromospheric reaction
to the flux emergence
in the photosphere
by using H$\alpha$ images
taken by
the Solar Magnetic Activity Research Telescope
(SMART)
at Hida Observatory.
In Section \ref{sec:observation},
we will introduce the observations
and the method of data reduction.
Analysis and the results will appear
in Section \ref{sec:results}.
Then, in Section \ref{sec:discussion},
we will discuss the observational results.
Finally, we will summarize the Paper
in Section \ref{sec:summary}.

\section{Observation and Data Reduction
  \label{sec:observation}}

In this Paper,
we studied NOAA AR 11081
formed in 2010 June,
in the northwest of the solar disk.
To measure the Doppler shift
and line-of-sight (LoS) magnetic field
in the photosphere,
we used Dopplergrams and magnetograms
taken by
SDO/HMI.
Also,
to study the chromospheric response
to the flux emergence,
we used H$\alpha$ images
taken by SMART at Hida Observatory.

\subsection{SDO/HMI Dopplergram and Magnetogram}

SDO/HMI continuously observes
the whole solar disk
at the 6173 \AA \ion{Fe}{1} line,
which is resolved by $4096^{2}$ pixels
\citep{sch12}.
To obtain the tracked data cubes
of the birth of AR 11081,
we used {\tt mtrack} module
\footnote{http://hmi.stanford.edu/teams/rings/mod\_mtrack.html}.
The data cubes
of the Doppler velocity
and the LoS magnetogram
have
a spatial resolution of $0.5\ {\rm arcsec}$
(1 pixel corresponds to $\sim 360\ {\rm km}$)
with $512^{2}$ pixel field-of-view (FoV),
and a temporal resolution of $45\ {\rm s}$
with a duration of $36\ {\rm hr}$,
starting at 12:00 UT
on 2010 June 10.
In the initial state,
the center of the $512^{2}$ FoV
is located at
N$22^{\circ}$ W$25.6^{\circ}$,
or ($+392, +383$) arcsecs
in solar disk coordinates.
Here, we applied Postel's projection,
that is, both Doppler and magnetic maps
are projected
as if seen from
directly above.
Then, to eliminate the effects of
the rotation of the Sun
and the orbital motion of the satellite,
and to determine the zero point
of the LoS velocity,
we reduced the mean velocity
from each Dopplergram.
Also, a 30-min (40-frame) moving average
was applied
to the Dopplergrams and magnetograms.

Figure \ref{fig:fov} is
the HMI magnetogram
of NOAA AR 11081
taken at 06:00 UT,
2010 June 11,
that is,
after the emergence started.
Here, white and black indicate
the positive and negative polarities,
respectively.
The diagonal line in this figure
is the slit
for the time-sliced diagram
in Section \ref{sec:slice}.
The slit angle
is chosen to fit
the first separating motion
of both polarities.
The square indicates
the region analyzed
in Section \ref{sec:histogram}
to measure the distributions
of the Doppler velocity
and the LoS
field strength.

\subsection{SMART H\boldmath{$\alpha$} Images}

SMART
at Hida Observatory,
Kyoto University,
consists of four different telescopes,
which are T1, T2, T3 and T4, respectively
\citep{uen04}.
They are placed on a tower
with a height of $16\ {\rm m}$.
T1 obtains H$\alpha$
full solar disk images
at high temporal and spatial resolution.
For studying the chromospheric reaction
to the photospheric flux emergence,
we analyzed the H$\alpha$ data
of 01:00--05:00 UT,
2010 June 11,
which resolves the full solar disk
with $4096^{2}$ pixels
(1 pixel corresponds
to $\sim 0.56\ {\rm arcsec}$)
and has a maximum temporal resolution
of 2 minutes.

In this study,
we only used H$\alpha$ line core images
(wavelength at $6562.8\ {\rm \AA}$).
First,
dark-current subtraction
and flat fielding
were performed
on the obtained SMART data.
Then, by taking
a cross-correlation
of the two consecutive images
to fix the position
of the target emerging active region,
we made a data cube
of H$\alpha$ images.
Note that H$\alpha$ image
is a simple zoom-up
of the full disk image,
while
Postel's projection is applied
to the HMI images.

\section{Data Analysis and Results
  \label{sec:results}}

Figure \ref{fig:evolution}
shows the temporal evolution
of the Dopplergram and the magnetogram
for 12 hours
from 18:00 UT,
2010 June 10.
In the Dopplergram,
the motion toward and away
from the observer are
shown in blue and red,
respectively.
At first,
during 18:00--00:00 UT,
the surface is relatively quiet
with some preceding magnetic elements
of both positive
and negative polarities.
An area 
with strong blue shift
($< -1\ {\rm km\ s}^{-1}$) appears
in the middle of the FoV
at 01:00 UT on 11 June,
which is gradually growing
in size.
After 3:00 UT,
the strong red shift
($> 1\ {\rm km\ s}^{-1}$) appears
and magnetic field emergence
takes place.
Both positive and negative polarities
move apart from each other.
Here, the separation
of the magnetic elements
is almost along the slit,
which is indicated as a diagonal line.
Finally, at 06:00 UT,
the red and blue areas
become faint.
The separated magnetic elements stop
and gather to form pores
at the boundary
of the emerging region.

In this section,
we first introduce the results
of time-slices
of the Dopplergrams and magnetograms
in Section \ref{sec:slice}.
Then, in Section \ref{sec:histogram},
we will clarify
the occurrence times
of the HDF
and the flux emergence,
and evaluate
the horizontal speed
of the HDF.
Section \ref{sec:chromosphere}
is dedicated
to showing
the chromospheric studies.

\subsection{Time-sliced Diagram\label{sec:slice}}

To examine the motion
of the magnetic elements
of positive and negative polarities
and the corresponding LoS velocity,
we made time-sliced diagrams
of HMI Dopplergrams and magnetograms.
The spatial slit is indicated
as a diagonal line
in Figure \ref{fig:fov}
and Figure \ref{fig:evolution},
which is placed
parallel to the separation
of both polarities.

Figure \ref{fig:slice}
is the time-sliced diagram
of the Dopplergram and the magnetogram
along the slit.
From the time-slice
of the magnetogram,
Figure \ref{fig:slice}(b),
we can see that
both positive and negative polarities
move apart from each other
from around 03:00 UT on June 11.
The speed of each element
is estimated to be
$\sim 1.2\ {\rm km\ s}^{-1}$,
which then drops
to $\sim 0.4\ {\rm km\ s}^{-1}$.
Thus, the separation
speed is $0.8$--$2.4\ {\rm km\ s}^{-1}$.
This deceleration
of the separated polarities
may reflect that
the polarities are reaching
the boundary
of the active region.
These elements then gathered
to create stronger pores,
of which the absolute LoS
field intensity is
greater than $200\ {\rm G}$.
One would find that
weak and small elements
of both polarities appear
between the main separating pores
during 03:00--09:00 UT
on June 11.
Also, the main positive pore
collides with
the preexisting negative polarity,
and they cancel
each other out.

In the Doppler slice,
Figure \ref{fig:slice}(a),
a pair of red and blue patterns
emerged at around 02:00 UT, June 11,
slightly earlier than
the appearance of the magnetic elements
in Figure \ref{fig:slice}(b).
The red and blue shift patterns
immediately started to separate,
and the propagation speed
of the patterns
(the slope of the patterns)
is about $0.4\ {\rm km\ s}^{-1}$.
Here,
we note that
the blue (red) pattern is located
disk-centerward (limbward),
which indicates that
the flow is divergent.
Moreover,
from the fact
that the divergent outflow
came before
the flux emergence,
we can assume that
the outflow
during this period
is caused by the plasma
escaping from
the rising magnetic flux.
It should be noted that
the trend
of the Doppler pattern
coming before the flux emergence
does not change
when we vary the thickness
of the slit.

However,
the determination
of the appearance time
of the Doppler pattern
associated with
the flux emergence
is difficult,
because the Doppler pattern,
especially the blue shift,
appeared at the location
where the supergranulation
showed blue shift
(21:00--01:00 UT).
The definition
of the flux emergence
and the appearance of the related Doppler pattern
is dealt with
in the next subsection
(\S \ref{sec:histogram}).

\subsection{Appearance times
  of the HDF and the flux emergence,
  and the velocity of the HDF
  \label{sec:histogram}}

It is not easy to determine
the timings
of the appearance of
the HDF
and the associated
flux emergence
from Figures \ref{fig:evolution}
and \ref{fig:slice}.
In particular,
we have to distinguish
the outflow
related to the flux emergence
from the preexisting
convective motions
of the quiet Sun
(e.g., granulations and supergranulations).
To clarify with significance
when the HDF
occurred
and when the magnetic flux emerged,
we studied the temporal changes
of the Doppler and magnetic patterns
from those before the emergence,
namely, patterns of the quiet Sun.
Also, in this subsection,
we describe how we evaluate
the horizontal speed
of the HDF.

\begin{figure}
  \includegraphics[scale=1.,clip]{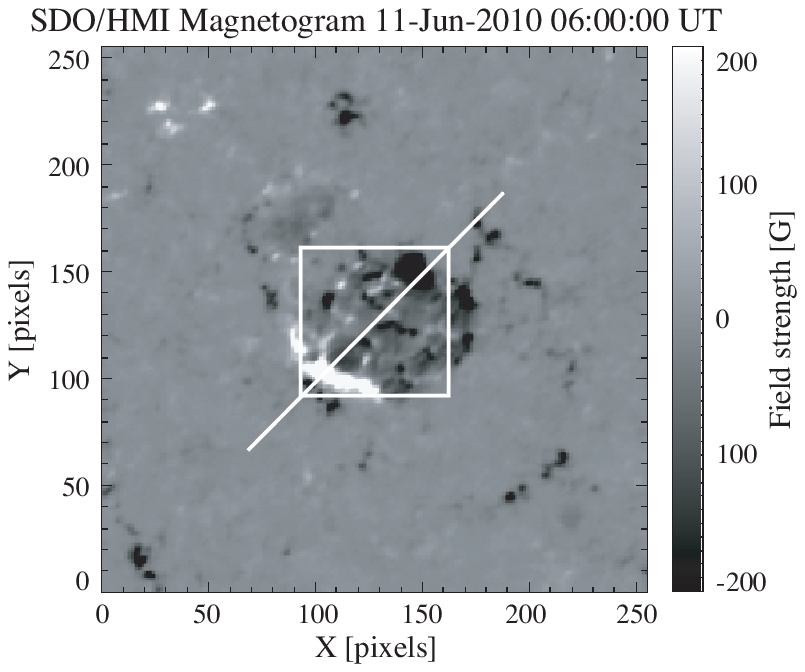}
  \caption{SDO/HMI magnetogram
    of NOAA AR 11081
    taken at 06:00 UT
    on 2010 June 11.
    Positive and negative polarities
    are indicated
    by white and black,
    respectively.
    One pixel
    corresponds to
    $\sim 350\ {\rm km}$.
    The diagonal line
    is the slit
    for time-sliced diagram
    (see Section \ref{sec:slice}).
    The square indicates the field
    in which temporal evolution
    of the Doppler velocity
    and the magnetic field strength
    (see Section \ref{sec:histogram})
    are analyzed.}
  \label{fig:fov}
\end{figure}

First, we plotted the histograms
of the Doppler velocity
and the absolute LoS
field strength
inside the square
of Figure \ref{fig:fov}
for each frame.
The size of the square
is $70\times 70$ pixels
$(\sim 25\times 25\ {\rm Mm}^{2})$,
which is selected
to include the emergence region.
As for the Dopplergram,
the apex of the histogram
was shifted
to fit the zero point.
Then, considering the photospheric condition
in the 3 hours
from 21:00 UT of June 10
to be sufficiently quiet,
we averaged up each 240 histograms
of the Dopplergrams and the magnetograms
in this period,
and regarded these averages
as reference quiet-Sun profiles.

In the left column
of Figure \ref{fig:histogram},
we show histograms
of the Doppler velocity
at five different times of June 11,
plotted over the reference
quiet-Sun profile.
Here we note that
the quiet-Sun profile
obtained is similar
to a Gaussian distribution.
The shade indicates
the standard deviation
above and below the reference.
As time goes by,
the profile becomes deviated
from the reference,
because the number of pixels
of which the absolute Doppler velocity
is greater
than $0.5\ {\rm km\ s}^{-1}$
increases.
The right column
of Figure \ref{fig:histogram}
is the residual
of the Doppler histogram
from the reference.
One standard deviation
is also shown as a shaded area.
At first,
the residual is below
one standard deviation level
for most of the velocity range.
From 02:00 UT, however,
the residual exceeds the deviation.

Figure \ref{fig:histogram_mag}
is the same as Figure \ref{fig:histogram},
but for the absolute
field strength
of the LoS magnetograms.
Here, the quiet-Sun profile
consists of a distribution with
a width of $\sim 10\ {\rm G}$
(about the precision
of the HMI magnetogram)
and some preexisting pores
within the FoV.
Thus, the profile is
different from
a Gaussian distribution.
The residual in the range
of $> 200\ {\rm G}$
further exceeds
one standard deviation level
from 04:00 UT.
After this time,
the residual of $> 200\ {\rm G}$
becomes well over the standard deviation,
because more and more flux is emerged
and stronger pores are created.

For the significance
of the measurement,
we define
the start time of the HDF
and the flux emergence
as the time
when each residual
of the Dopplergrams and the magnetograms
exceeded one standard deviation level.
To know these times,
we show in Figure \ref{fig:timing}
each time-evolution
of the residuals
(taken from and averaged over
the range
$[-0.8\ {\rm km\ s}^{-1}, -0.4\ {\rm km\ s}^{-1}]$
and $[0.4\ {\rm km\ s}^{-1}, 0.8\ {\rm km\ s}^{-1}]$
for Dopplergram,
and the range $[200\ {\rm G}, 300\ {\rm G}]$
for magnetogram),
plotted over
one standard deviation.
In this figure,
the residual of the Dopplergram
becomes over the standard deviation
at 01:23 UT on 11 June,
while that of the magnetogram
exceeds the level
at 03:06 UT.
That is,
the appearance of the HDF
came before the flux emergence
by about 100 minutes.

During this period,
it is expected that
the flow is mainly horizontal
and a vertical component
is less dominant.
Thus, we can calculate
the horizontal velocity
from the residual distribution
of the Doppler velocity
(Figure \ref{fig:histogram}),
by considering
the geometric effect.
The relation between
the horizontal velocity $V_{\rm h}$
and the Doppler velocity $V_{\rm D}$ is
$V_{\rm h}=V_{\rm D}/\sin{\theta}$,
where $\theta$ is the heliocentric angle
of the emerging region
measured from the disk center.
From 01:23 to 03:06 UT,
the Doppler velocity range
where the residual exceeds
the one standard deviation
is typically
0.4--$1.0\ {\rm km\ s}^{-1}$,
which is up to
$1.5\ {\rm km\ s}^{-1}$,
and the heliocentric angle is
$\sim 40^{\circ}$.
Therefore,
the horizontal velocity
is calculated to be
$0.6$--$1.5\ {\rm km\ s}^{-1}$,
and the maximum is
$2.3\ {\rm km\ s}^{-1}$.

Here, we comment
on the selection
of the field-strength range
($[200\ {\rm G}, 300\ {\rm G}]$)
and its dependence
on the start time
of the flux emergence.
If we use the lower strength range,
for example $[50\ {\rm G}, 100\ {\rm G}]$
or $[100\ {\rm G}, 200\ {\rm G}]$,
at which the residual exceeds
one standard deviation level faster
(Figure \ref{fig:histogram_mag}, right column),
the start time of the flux emergence
is calculated to be much earlier.
In the present analysis,
however,
the strength range
$[200\ {\rm G}, 300\ {\rm G}]$
is used,
since the number of the pixels of $>200\ {\rm G}$
is so small in the quiet Sun
that the flux emergence is easily detected
when it occurs.
We confirmed this fact
by applying the same analysis
on the quiet-Sun data.
As for the dependence of the strength range
on the observation results,
we tested the analysis
with various ranges,
which is summarized in
Table \ref{tab:range}.
From this table
one can see that
the start time does not so change
for [$200\ {\rm G}, 300\ {\rm G}$],
[$300\ {\rm G}, 400\ {\rm G}$],
and [$400\ {\rm G}, 500\ {\rm G}$] cases.

We also checked the dependence
of the size of the square
where the histograms are made
(Fig. \ref{fig:fov}),
which is summarized in
Table \ref{tab:size}.
Here, the time difference
is almost constant
for various square sizes
and is about 100 min.
With increasing square size,
the ratio of high-speed or strong pixels
in the square reduces.
At the same time,
the quiet-Sun reference profile
becomes more accurate
and one standard deviation level decreases.
Therefore, in total,
the time difference remains constant.

\subsection{Chromospheric Response
  \label{sec:chromosphere}}

In this subsection,
we investigate
the time-evolution
of the H$\alpha$ intensity
to examine the relation
between the chromosphere
and the photosphere
in this studied event.
Figure \ref{fig:ha}(a)
is a sample image
of the SMART H$\alpha$ data.
The color and contours
indicate
the relative H$\alpha$ intensity.
In this figure,
there are two bright regions
(plages)
in the middle of the FoV.
Then,
along the slit
of Figure \ref{fig:ha}(a),
we made a time-sliced diagram
for 4 hours
starting at 01:00 UT, 11 June,
which is shown
as Figure \ref{fig:ha}(b).
Note that the slit
in Figure \ref{fig:ha}(a)
is not exactly the same as
that in Figure \ref{fig:fov},
since the H$\alpha$ data
is a simple closeup view
of the full disk image,
while Postel's projection
is applied to the HMI data.
Thus,
from this study,
we can only determine
the appearance time
of the chromospheric brightening.

In Figure \ref{fig:ha}(b),
the first bright source
at the slit location
of $5\times 10^{4}\ {\rm km}$
starts at 02:40 UT.
However, it was found that
this brightening
is due to the activity
among the preexisting quiet-Sun pores
of both polarities,
which later collide with
positive patches
of the newly emerging flux
(see Section \ref{sec:slice}).
It is difficult to
separate this bright source
into activity
of the preexisting pores
and that of the newly emerged
positive pores.
The second source
located at $7\times 10^{4}\ {\rm km}$
starts at 03:20 UT,
and there was
no preceding pore
in this region.
Therefore,
we consider that
the second source
is entirely due to
the newly
emerged negative pores,
and determine that
the chromospheric reaction
starts at this time
(03:20 UT;
indicated by a dashed line
in Figure \ref{fig:ha}(b)).
The two chromospheric sources
are located
just over the positive
and negative polarities
in the photosphere.

\section{Discussion\label{sec:discussion}}

\subsection{Mechanism of the Time Difference
  \label{sec:mechanism}}

In this Paper
we analyze
the newly emerging active region
and find that
there is a time difference
between the appearance of
the horizontal divergent flow (HDF)
and the corresponding flux emergence;
the HDF
appears prior to the flux emergence
by about 100 minutes.

According to the thin-flux-tube
simulation \citep{fan09},
the rising speed of the flux tube
accelerates
from the top few tens of Mm
of the convection zone.
However, at the same time,
the flux tube expands
as the external density (pressure) 
decreases with height.
The radius of the tube eventually exceeds
the local pressure scale height
at a depth of $\sim 20\ {\rm Mm}$
and the thin-flux-tube approximation
breaks out.
Recently, our numerical simulations
using the fully compressed MHD,
including the convection zone,
the photosphere,
and the corona
in a single computational box,
have revealed that
the rising flux tube
decelerates
in the uppermost convection zone
\citep{tor11,tor12}.
It is because
the plasma on the flux tube piles up
between the apex of the tube
and the subadiabatically stratified photosphere ahead,
and the plasma inhibits the rising motion of the flux tube.
Then, the accumulated plasma
in turn extends the tube laterally.
This accumulation becomes effective
from the depth
where the apex of the tube
becomes ``flat''.
This critical depth
is also considered as
being
where the tube's radius exceeds
the local pressure scale height
(depth $\sim -20\ {\rm Mm}$).
The lateral expansion
of the flux tube
appears
similar to those
found by \citet{mag01} and \citet{arc04}.
However, their expansions occur
because the tubes themselves
move into the subadiabatic photosphere.

As the rising tube approaches
the photosphere,
the accumulated plasma
on the rising tube
escapes horizontally
around the surface
and is observed
as an HDF,
while the tube
stops beneath the surface.
Since the flux is
continuously transported
from below,
the magnetic pressure gradient
at the photosphere
enhances
and the further emergence
to the upper atmosphere
starts
due to the magnetic buoyancy instability.
When the flux resumes rising,
it can be observed as a ``flux emergence''
at the photospheric level.
Therefore,
the time difference
detected in this Paper implies
the period of latency
during which
the flux tube
reaching the photosphere
develops the magnetic buoyancy instability.
The growth time
of the instability is,
however,
complicated
and may be related
to many parameters
of the rising flux tube
such as field strength,
total flux, twist, etc.
Thus, we shall leave
the estimation of the time gap
for our future numerical research.

\subsection{Depth of the Magnetic Flux
  \label{sec:model}}

To describe the relation
between the HDF
and the contributing upflow
below the surface,
we make a simple model,
which is schematically illustrated
as Figure \ref{fig:model}.
When the magnetic flux tube has emerged
from the deeper convection zone,
an upflow region is formed
in front of the flux tube.
If the typical size
of this region is $L$
and the velocity is $V_{\rm up}$,
the mass flux passing through
the area of $\pi L^{2}$
can be described as
\begin{eqnarray}
  F_{1}=\rho_{1} V_{\rm up} \pi L^{2},
  \label{eq:f1}
\end{eqnarray}
where $\rho_{1}$ is the plasma density.
Next, the photospheric plasma
that escapes from the upflow
propagates the surface
as an HDF.
If we write the horizontal velocity
at the radial distance $r$
as $V_{\rm h}(r)$,
the thickness as $T$,
and the density as $\rho_{2}$,
the mass flux passing through $2\pi rT$ is
\begin{eqnarray}
  F_{2}=2\pi r \rho_{2}TV_{\rm h}(r).
  \label{eq:f2}
\end{eqnarray}
These fluxes,
$F_{1}$ and $F_{2}$,
are assumed to be conserved.
Therefore,
from Equations (\ref{eq:f1}) and (\ref{eq:f2}),
the upflow velocity is
\begin{eqnarray}
  V_{\rm up}=\frac{2\rho_{2}}{\rho_{1}}\frac{rTV_{\rm h}(r)}{L^{2}}.
  \label{eq:vup1}
\end{eqnarray}

As a result of the observational study,
the horizontal speed is
$V_{\rm h}\sim 1\ {\rm km\ s}^{-1}$
at $r=5000\ {\rm km}$.
Here we assume that
(a) plasma density is almost
uniform
around the photosphere,
i.e., $\rho_{1}\sim \rho_{2}$,
(b) the thickness is about
the local pressure scale height,
$T\sim 200\ {\rm km}$,
and (c) the size of the upflow
is $4000\ {\rm km}$
(the smallest distance
between the blue and red patterns
in Figure \ref{fig:slice});
$L\sim 2000\ {\rm km}$.
Under these assumptions,
Equation (\ref{eq:vup1}) reduces to
$V_{\rm up}=0.5\ {\rm km\ s}^{-1}$.
The time gap
between the HDF
appearance
and the flux emergence
was observed to be $100\ {\rm min}$.
Therefore,
the depth
that the apex of the magnetic flux
transited across
after it decelerated,
is estimated to be
$\sim 3000\ {\rm km}$,
if the flux tube rises
at the same rate
as the upflow.

\begin{figure}
  \includegraphics[scale=0.8,clip]{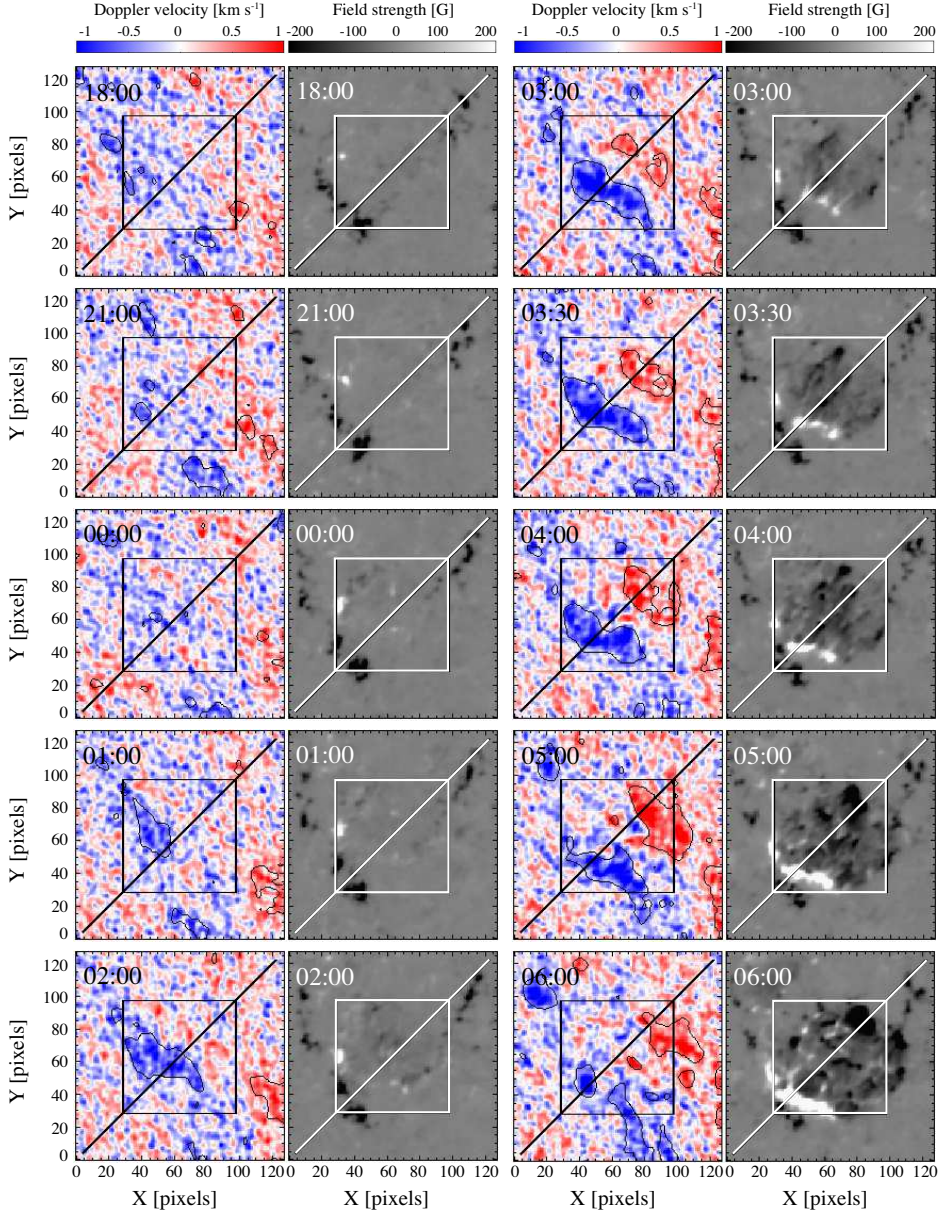}
  \caption{Temporal evolutions
    of the Dopplergram (red-blue map)
    and the magnetogram (white-black map)
    for 12 hours
    from 18:00 UT,
    2010 June 10
    to 06:00 UT,
    June 11.
    The diagonal line
    and the square
    in each panel
    are the slit for time-sliced diagrams
    (Section \ref{sec:slice})
    and the field-of-view
    in which histograms are made
    (Section \ref{sec:histogram}),
    respectively.
    In the Doppler maps,
    the motion toward and
    away from the observer
    are indicated in blue and red,
    respectively,
    and contours indicate
    the smoothed isolines
    of $\pm 1\ {\rm km\ s}^{-1}$.
    In the magnetograms,
    positive and negative polarities
    are shown with white and black,
    respectively.}
  \label{fig:evolution}
\end{figure}

\begin{figure}
  \includegraphics[scale=0.9,clip]{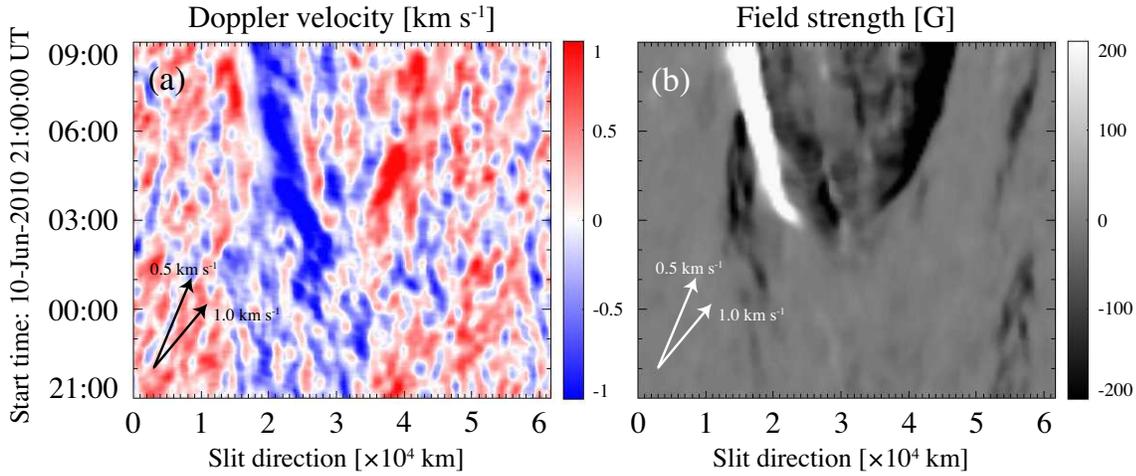}
  \caption{Time-slice diagram of
    (a) Dopplergram and (b) magnetogram
    along the slit
    shown in Figure \ref{fig:fov}
    and Figure \ref{fig:evolution}.
    Slit direction is selected limbward,
    i.e., the distance from the disk center
    increases with the horizontal axis.
    The duration of the time-slice
    is 12 hours,
    starting from 21:00 UT
    on 2010 June 10.
    In the Doppler time-slice (a),
    the motion toward (away from)
    the observer is indicated
    by blue (red) color.
    In the magnetogram time-slice (b),
    positive (negative) polarity
    is shown as white (black).
    Arrows give the apparent velocity.}
  \label{fig:slice}
\end{figure}

\begin{figure}
  \includegraphics[scale=0.9,clip]{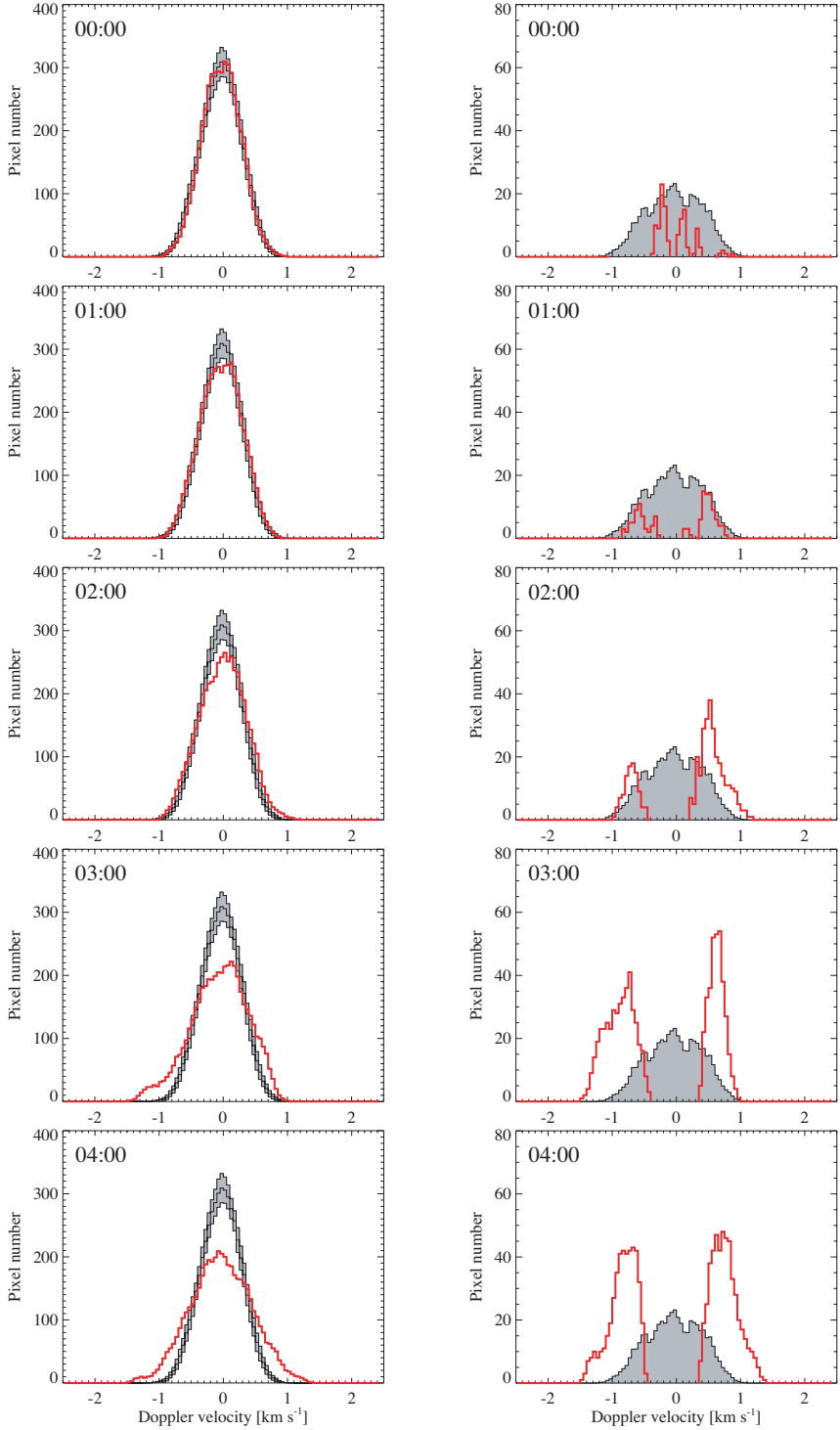}
  \caption{({\it Left}) The histogram
    of the Doppler maps
    at five different times
    of 2010 June 11,
    indicated by red line,
    plotted over the quiet-Sun reference profile
    (middle black line).
    Shade indicates
    the standard deviation
    above and below the reference.
    ({\it Right}) The residual
    of the histogram
    from the reference,
    indicated by red line.
    Shade is one standard deviation level.}
  \label{fig:histogram}
\end{figure}

\begin{figure}
  \includegraphics[scale=0.9,clip]{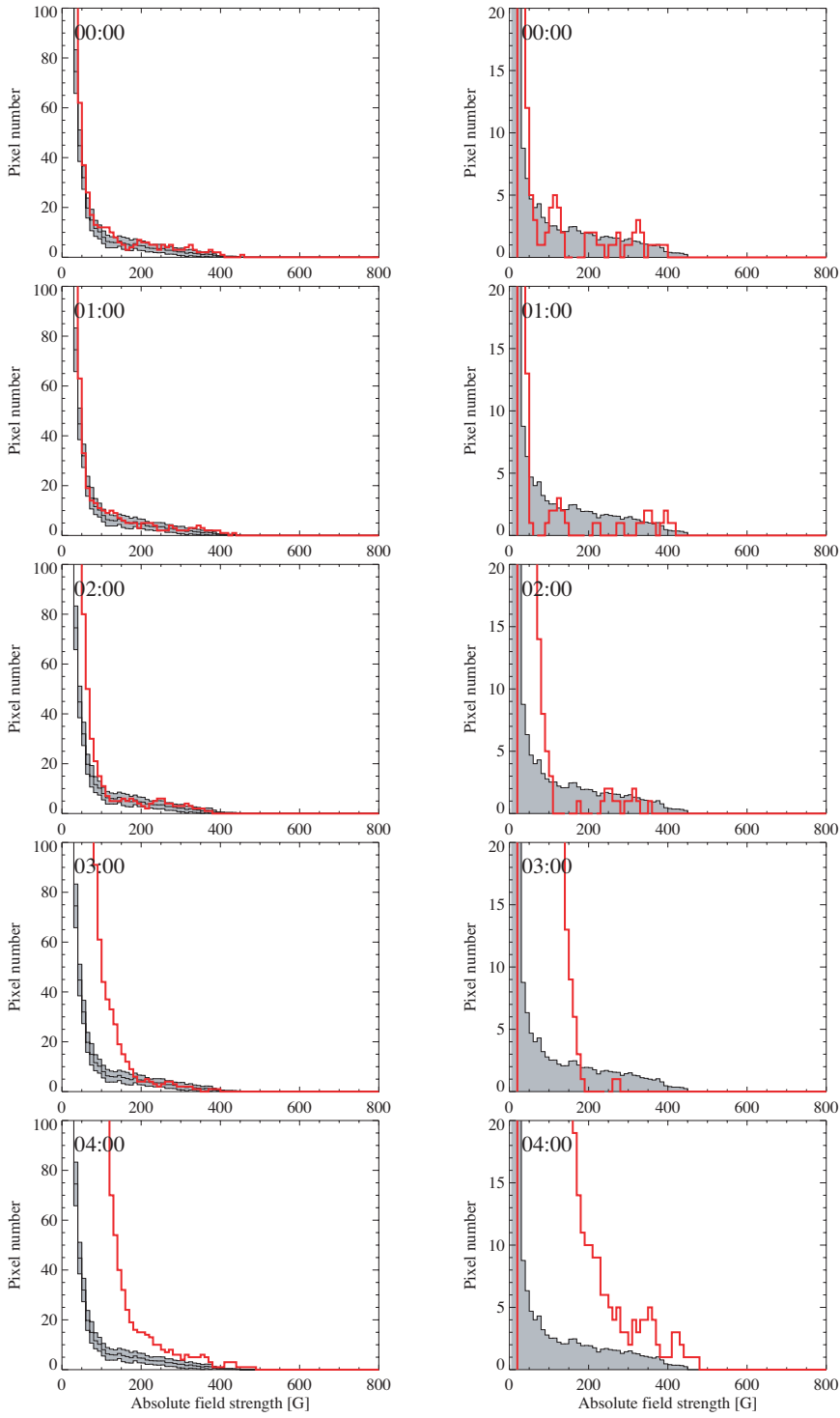}
  \caption{
    Same
    as Figure \ref{fig:histogram},
    but for the absolute
    field strength
    of the LoS magnetograms.}
  \label{fig:histogram_mag}
\end{figure}

\begin{figure}
  \includegraphics[scale=1.,clip]{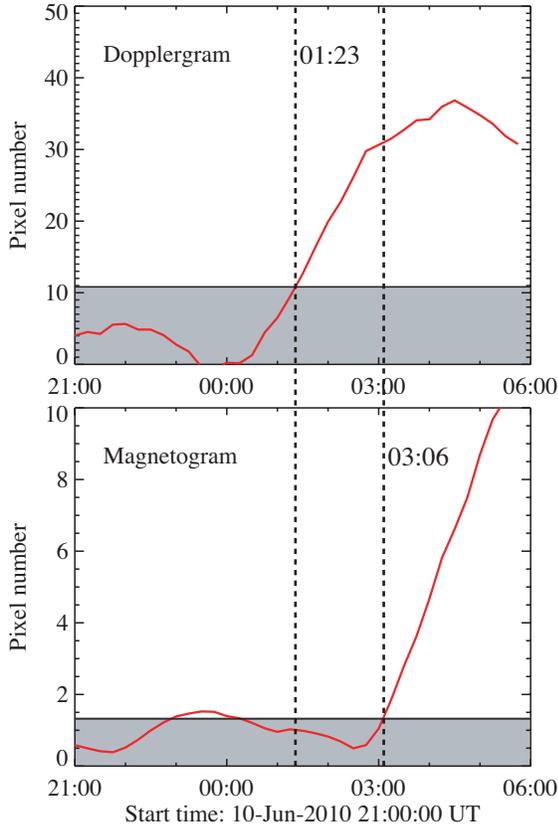}
  \caption{({\it Top}) Time-evolution
    of the residual
    of the Doppler histogram
    (Figure \ref{fig:histogram}),
    averaged over the range of
    $[-0.8\ {\rm km\ s}^{-1}, -0.4\ {\rm km\ s}^{-1}]$
    and $[0.4\ {\rm km\ s}^{-1}, 0.8\ {\rm km\ s}^{-1}]$,
    starting from
    21:00 UT,
    2010 June 10.
    The shaded area is
    one standard deviation level.
    The residual exceeds
    the standard deviation level
    at 01:23 UT
    on June 11.
    ({\it Bottom})
    The same for the magnetogram
    (Figure \ref{fig:histogram_mag})
    over the range of
    $[200\ {\rm G}, 300\ {\rm G}]$.
    The residual exceeds
    at 03:06 UT.}
  \label{fig:timing}
\end{figure}

\begin{figure}
  \includegraphics[scale=1,clip]{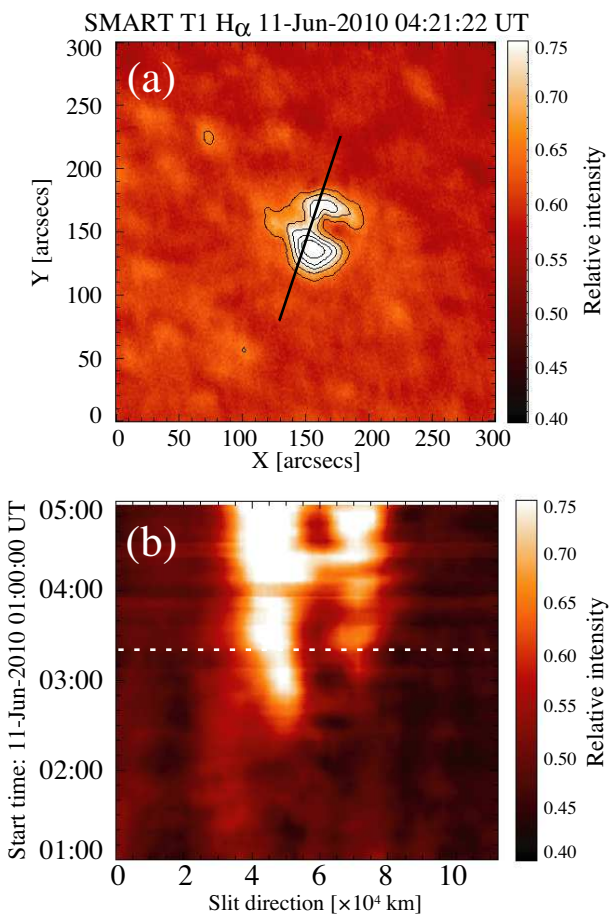}
  \caption{(a) A sample image
    of SMART chromospheric H$\alpha$ data.
    The color and contours
    indicate the relative H$\alpha$ intensity.
    Contour levels are
    0.65, 0.70, 0.75, 0.80, and 0.85,
    respectively.
    The slit
    in the middle of the panel
    is used to
    make a time-sliced diagram.
    (b) Time-sliced diagram
    of H$\alpha$ image
    for 4 hours
    from 01:00 UT,
    2010 June 11.
    Color is the relative intensity.
    Slit direction (horizontal axis)
    starts from the bottom left
    of the slit in Panel (a).
    The dashed line
    indicates the time 03:20 UT
    (see text for details).}
  \label{fig:ha}
\end{figure}

\begin{figure}
  \includegraphics[scale=1,clip]{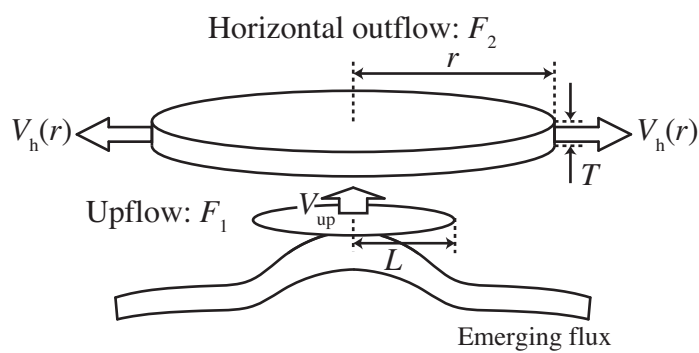}
  \caption{Schematic illustration
    of the model of flux emergence
    and the divergent horizontal flow.
    When the magnetic flux is emerged
    from the deeper convection zone,
    an upflow region
    with a size of $L$
    and a velocity of $V_{\rm up}$
    is formed
    above the magnetic flux.
    At the photospheric level,
    the plasma is pushed away
    to create a divergent horizontal flow
    with a thickness of $T$.
    Here, $r$ is the radial distance
    from the center of the emerging region
    and $V_{\rm h}(r)$
    is the horizontal velocity
    at $r$.
    $F_{1}$ and $F_{2}$ are
    the mass fluxes
    of the upflow
    and the horizontal outflow,
    respectively.}
  \label{fig:model}
\end{figure}

In this section,
for simplicity,
we assumed that
the apex of the rising flux is circular,
and that the outflow velocity $V_{\rm h}$
is only a function of $r$.
From Figure \ref{fig:evolution},
however,
it seems that the HDF is not axisymmetric
and is stronger
in the direction of flux emergence
(the northwest-southeast slit in this figure).
This property is consistent
with our preceding numerical results;
the photospheric plasma flow
is found to be
along the direction of flux emergence
\citep[see][Fig. 4]{tor12}.
Moreover,
in that simulation,
the twist of the rising flux tube
is stronger
and the magnetic field
at the tube's surface
is almost perpendicular
to the axis of the tube.
In the later phase of
the target AR of this Paper,
the separation of
positive and negative polarities
shifted into the northeast-southwest direction,
i.e., perpendicular to the diagonal line
in Figure \ref{fig:evolution}.
Taking into account
the previous numerical results,
and considering that
the observed NE-SW direction indicates
the axis of the flux tube
that forms this AR,
we can think that the twist
of this flux tube is tight,
and therefore the flow
is in the NW-SE direction.

\subsection{Relations with Recent Observations:
  HDF as a precursor
  \label{sec:seismology}}

Using SOHO/MDI,
\citet{gri07} observed NOAA AR 10488
and found that
upflows of matter
with a high velocity
($\gtrsim 0.4\ {\rm km\ s}^{-1}$)
preceded flux emergences
by 8 and 13 min.
Thus,
the last $\sim 10$ min
of the divergent Doppler pattern
observed in our study
that remained for 100 min,
may contain
the upward motion.
However,
for most of the period,
the flow is expected
to remain horizontal.
Note that
the upflow velocity of
$\gtrsim 0.4\ {\rm km\ s}^{-1}$
reported
by \citet{gri07}
may be the speed
of a magnetic flux
rising in the photosphere.
As for the estimated velocity
($V_{\rm up}=0.5\ {\rm km\ s}^{-1}$)
in Section \ref{sec:model},
this value indicates
the emergence speed
of a magnetic flux
in the uppermost convection zone.

By means of time-distance helioseismology,
\citet{ilo11} detected
strong acoustic travel-time anomalies
as deep as 65 Mm,
1 to 2 days
before the flux rate reaches its peak,
and (in most cases)
a few hours before
the start of
the flux appearance
at the surface
\citep[see also][]{kos08,kos09}.
These anomalies are
considered as
signs of the rising
magnetic flux.
Taking account
of our numerical simulations
\citep[e.g.][]{tor12},
it is consistent
to interpret
this helioseismic anomaly
as a result
of the effect
similar to the plasma accumulation;
external media
may be perturbed or compressed
by the rising motion
of the magnetic flux.
The importance
of the helioseismic anomaly
in \citet{ilo11}
and the HDF in our study
is that
these phenomena occur
prior to the flux emergence
at the photosphere.
That is,
these are
the precursors
of the flux emergence.
By combining two types
of observations,
sunspot appearances
may be predicted
in the near future.

\subsection{Further Emergence to the Upper Atmosphere
  \label{sec:further}}

In Section \ref{sec:chromosphere},
we found that the H$\alpha$ brightenings
(plages) were located
over the positive and negative pores
in the photosphere.
This indicates that
the brightenings
are caused by the plasma
flowing down along magnetic loops
that connect the photospheric magnetic elements
\citep[see][Figure 10]{shi89}.
The appearance of the chromospheric source
was at 03:20 UT
on June 11,
while the flux emergence
was at 03:06 UT.
If we assume the H$\alpha$ formation height
as $2000\ {\rm km}$,
the rise velocity of the magnetic field is 
$\sim 2.5\ {\rm km\ s}^{-1}$.
This value is smaller than
the observed speed
of the chromospheric arch filament system (AFS)
of $\sim 20\ {\rm km\ s}^{-1}$
\citep[e.g.][]{bru67},
which implies that
the actual rise speed
is faster than $2.5\ {\rm km\ s}^{-1}$
and it takes some time
to create H$\alpha$ plage
after the flux reaches
the chromospheric height.

\section{Summary\label{sec:summary}}

In this Paper,
we have observed
the horizontal divergent flow (HDF)
prior to the flux emergence
by using SDO/HMI Dopplergram and magnetogram.
The presence of the HDF
was predicted
by our preceding numerical simulations
\citep[e.g.][]{tor12}.
The HMI's continuous observation
of the whole solar disk provides
the means to analyze
the earlier stage
of the flux emergence.
The summary of the observation
is given
as Table \ref{tab:summary}.

First, we made time-slices of
Dopplergram and LoS magnetogram
of NOAA AR 11081.
From the magnetic slice,
we found that
the magnetic elements
of positive and negative polarities
separated from each other.
The apparent speed
of a single element was,
at first, $1.2\ {\rm km\ s}^{-1}$.
The speed then dropped
to $0.4\ {\rm km\ s}^{-1}$
and the elements gathered
to create stronger pores
of $>200\ {\rm G}$.
In the Doppler slice,
a pair of blue and red pattern
was observed to separate,
slightly earlier than
the flux emergence,
and the blue (red) pattern
was located disk-centerward (limbward).
This indicates that
the HDF
appeared prior to the flux emergence.
According to our previous numerical experiments,
the outflow is mainly horizontal
during the period
from the appearance of the outflow
to the emergence of the magnetic flux.

Secondly,
we evaluated the times of the HDF
appearance
and the flux emergence.
To determine these times
with significance,
we studied the temporal changes
of the Doppler and magnetic patterns
from those of the quiet Sun,
and defined them as
the times when each profile exceeded
one standard deviation
of its quiet-Sun profile.
As a result,
the Doppler profile was found to
deviate from the quiet-Sun profile
at 01:23 UT, 2010 June 11,
while the magnetic profile
deviated at 03:06 UT.
Therefore,
the time difference was
about 100 minutes.
Also, by considering the heliocentric angle,
the horizontal speed of
the HDF in this time gap
was estimated to be
$0.6$--$1.5\ {\rm km\ s}^{-1}$,
up to $2.3\ {\rm km\ s}^{-1}$.

The creation of the HDF
is due to the density accumulated
on the apex of the flux tube
during its ascent
in the convection zone.
This accumulation occurs
between the flattened apex
of the rising flux tube
and the subadiabatically stratified photosphere.
The compressed plasma
escapes horizontally
around the photosphere,
which was observed
in this Paper.
After the magnetic flux
is sufficiently intensified,
the magnetic buoyancy instability
is triggered
and the magnetic field restarts
into the upper atmosphere,
which was also seen
as a flux emergence
in this Paper.
Therefore, the time difference
of $\sim 100$ min
may reflect
the latency
during which
the flux is waiting
for the instability onset. 

Applying a simple model
of the horizontal flow
and the corresponding upflow
beneath the surface,
we speculated that
the depth of the magnetic flux
is about $3000\ {\rm km}$.
Previously,
SOHO/MDI found that
an upflow preceded the flux emergence
by about 10 minutes
\citep{gri07}.
This implies that the last
$\sim 10$ min of the divergent outflow
may include the upward motion.
Even so,
for most of the period,
the outflow remains horizontal.

Moreover,
using H$\alpha$ images
taken by SMART,
we studied chromospheric response
to the flux emergence
at the photosphere.
The time-slice
showed a pair of
H$\alpha$ plages,
which started from 03:20 UT,
that is,
$\sim 14$ min
after the flux emergence.
The location of these brightenings
were just over the photospheric pores.
Therefore,
we speculated that
these brightenings are caused by
the plasma precipitating along
the magnetic fields
that connect photospheric pores
of both polarities.

The time gap
between the HDF occurrence
and the flux emergence
will be investigated
in our future numerical study.
As for the observational study,
the statistical analysis on HDFs
would be the next target.
Another importance
of observing HDF is that
this phenomenon
can be considered as a precursor,
which may allow us
to predict sunspot formation
that occurs in several hours.



\acknowledgments

We thank the SDO/HMI team
for data support
and useful discussions.
S.T. thanks Dr. A. Kosovichev
for arranging his stay
at Stanford University.
This work was supported
by the JSPS Institutional Program
for Young Researcher Overseas Visits,
and by the Grant-in-Aid
for JSPS Fellows.
We are grateful
to the GCOE program instructors
of the University of Tokyo
for proofreading/editing assistance.
We also appreciate
the thorough and helpful comments
by the anonymous referee.

\clearpage



\clearpage

\clearpage

\clearpage

\clearpage

\clearpage

\clearpage

\clearpage

\clearpage

\begin{deluxetable}{cc}
  \tablecaption{Dependence of the field-strength range
    \label{tab:range}}
  \tablewidth{0pt}
  \tablehead{
    \colhead{Field-strength range [G]} &
    \colhead{Start of flux emergence}
   }
   \startdata
   0--50 & -:- UT\tablenotemark{a} \\
   50--100 & 01:25 UT \\
   100--200 & 02:15 UT \\
   200--300\tablenotemark{b} & 03:06 UT \\
   300--400 & 03:20 UT \\
   400--500 & 03:25 UT \\
   500--600 & -:- UT\tablenotemark{c} \\
   \enddata
   \tablenotetext{a}{Residual is always below one standard deviation level}
   \tablenotetext{b}{Used in this Paper}
   \tablenotetext{c}{One standard deviation level is not defined}
\end{deluxetable}

\clearpage

\begin{deluxetable}{cccc}
  \tablecaption{Dependence of the square size
    \label{tab:size}}
  \tablewidth{0pt}
  \tablehead{
    \colhead{Square size [pixels]} &
    \colhead{HDF appearance} &
    \colhead{Start of flux emergence} &
    \colhead{Time difference [min]}
   }
   \startdata
   $50\times 50$ & 01:00 UT & 02:35 UT & 95 \\
   $60\times 60$ & 01:25 UT & 03:00 UT & 115  \\
   $70\times 70$\tablenotemark{a} & 01:23 UT & 03:06 UT & 103 \\
   $80\times 80$ & 01:35 UT & 03:15 UT & 100 \\
   $90\times 90$ & 01:25 UT & 03:20 UT & 115 \\
   $100\times 100$ & 01:35 UT & 03:20 UT & 105 \\
   $110\times 110$ & 01:45 UT & 03:05 UT & 80 \\
   $120\times 120$ & 01:50 UT & 03:05 UT & 75 \\
   \enddata
   \tablenotetext{a}{Used in this Paper}
\end{deluxetable}

\clearpage

\begin{deluxetable}{cc}
  \tablecaption{Summary of the AR 11081 observation
    \label{tab:summary}}
  \tablewidth{0pt}
  \tablehead{
    \colhead{Physical value} & \colhead{Observational results}
   }
   \startdata
   Field strength & $< 500\ {\rm G}$ \\
   Unsigned total flux & $\sim 10^{21}\ {\rm Mx}$ \\
   Region area & $1.2\times 10^{9}\ {\rm km}^{2}$ \\
   Appearance of horizontal outflow & 01:23 UT\tablenotemark{a} \\
   Start of flux emergence & 03:06 UT\tablenotemark{a} \\
   Start of chromospheric response & 03:20 UT\tablenotemark{a} \\
   Time difference\tablenotemark{b} & $\sim 100\ {\rm min}$ \\
   Apparent speed\tablenotemark{c} &
    $1.2 \rightarrow 0.4\ {\rm km\ s}^{-1} $ \\
   Horizontal velocity\tablenotemark{d} &
    $0.6$--$1.5\ {\rm km\ s}^{-1}$ \\
    & (max $2.3\ {\rm km\ s}^{-1}$) \\
   \enddata
   \tablenotetext{a}{On 11 June, 2010}
   \tablenotetext{b}{Time difference between
     the appearance of the horizontal outflow
     and the flux emergence}
   \tablenotetext{c}{Apparent speed
     of the magnetic elements}
   \tablenotetext{d}{Horizontal velocity
     of the surface plasma
     prior to the flux emergence}
\end{deluxetable}




\end{document}